\PassOptionsToPackage{table}{xcolor}
\documentclass{aastex631}
\usepackage{amsmath,amssymb}

\shorttitle{Residual Energy and Broken Symmetry in RMHD}
\shortauthors{Dorfman et al.}

\begin{document}

\title{Residual Energy and Broken Symmetry in Reduced Magnetohydrodynamics}

\author{S.~Dorfman}
\affiliation{Space Science Institute, Boulder, Colorado, USA}
\affiliation{Department of Physics and Astronomy, University of California, Los Angeles, California, USA}
\author{M.~Abler}
\affiliation{Space Science Institute, Boulder, Colorado, USA}
\affiliation{Department of Physics and Astronomy, University of California, Los Angeles, California, USA}
\author{S.~Boldyrev}
\affiliation{Department of Physics, University of Wisconsin–Madison, Madison, Wisconsin, USA}
\affiliation{Space Science Institute, Boulder, Colorado, USA}
\author{C.~H.~K.~Chen}
\affiliation{Department of Physics and Astronomy, Queen Mary University of London, London, UK}
\author{S.~Greess}
\affiliation{Department of Physics and Astronomy, Queen Mary University of London, London, UK}



\begin{abstract}

Alfv{\'e}nic interactions which transfer energy from large to small spatial scales lie at the heart of magnetohydrodynamic turbulence.  An important feature of the turbulence is the generation of negative residual energy -- excess energy in magnetic fluctuations compared to velocity fluctuations. By contrast, an MHD Alfv{\'e}n wave has equal amounts of energy in fluctuations of each type. Alfv{\'e}nic quasimodes that do not satisfy the Alfv{\'e}n wave dispersion relation and exist only in the presence of a nonlinear term can contain either positive or negative residual energy, but until now an intuitive physical explanation for why negative residual energy is preferred has remained elusive.  This paper shows that the equations of reduced MHD are symmetric in that they have no intrinsic preference for one sign of the residual energy over the other.  An initial state that is not an exact solution to the equations can break this symmetry in a way that leads to net-negative residual energy generation.  Such a state leads to a solution with three distinct parts: nonresonant Alfv{\'e}nic quasimodes, normal modes produced to satisfy initial conditions, and resonant normal modes that grow in time.  The latter two parts strongly depend on initial conditions; the resulting symmetry breaking leads to net-negative residual energy both in Alfv{\'e}nic quasimodes and $\omega=k_\parallel{V_A}=0$ modes.  These modes have net-positive residual energy in the equivalent boundary value problem, suggesting that the initial value setup is a better match for solar wind turbulence.

\end{abstract}

\keywords{Magnetohydrodynamics(1964) --- Interplanetary turbulence(830) --- Alfven waves(23) --- Solar wind(1534)}


\section{Introduction}\label{sec:intro}

Turbulence is a fundamental and widespread aspect of plasma behavior, occurring in nearly every natural plasma system that we observe, from the large scales of galaxy clusters \citep{schuecker04, subramanian06} down to the plasmas within our solar system \citep{saur02} including the solar wind \citep{bruno13, chen16}.  Leading theories of magnetohydrodynamic turbulence (e.g., \citealp{goldreich95, boldyrev05}) describe counter-propagating Alfv{\'e}nic interactions that transfer energy from large to small spatial scales.  An important feature of the turbulence is the generation of negative residual energy -- excess energy in normalized magnetic fluctuations (${\bf \delta{b}}={\bf \delta{B}}/\sqrt{4\pi{n}{m_i}}$) compared to velocity fluctuations (${\bf \delta{v}}$) \citep{matthaeus82, bavassano98, chen13}.  By contrast, an MHD Alfv{\'e}n wave has equal amounts of energy in fluctuations of each type.  While solar wind turbulence measurements do show an outward flux of low frequency (below ion cyclotron), incompressible modes with highly correlated magnetic and velocity fluctuations \citep{coleman67, belcher71} (all characteristics consistent with linear Alfv{\'e}n waves), the presence of residual energy is taken to indicate a key role for nonlinear physics \citep{matthaeus21}.  In other words, fully developed turbulence retains only some of the properties of the relevant linear wave modes \citep{groselj19, verscharen19}.  However, there is to date no intuitive physical explanation why negative residual energy is preferentially generated, how this depends on physical parameters such as initial conditions, and how this relates to departures from a physical picture dominated by interacting normal mode waves.

Several ideas have been proposed over the years to explain residual energy in the solar wind.  Non-MHD corrections to the Alfv{\'e}n speed due to temperature anisotropies were put forward as a resolution \citep{belcher71, matthaeus82}, but it was later shown that these corrections cannot generate sufficient residual energy to fully account for observations \citep{bavassano00}.  Residual energy could also be due to coherent structures such as flux tubes and current sheets that either arise from or develop separately from the turbulence \citep{matthaeus86, bowen18a}.  Conversion to compressible fluctuations \citep{ofman98} or fluctuations generated by other instabilities (e.g.~mirror and firehose \citep{hellinger06}) could alter the fundamental nature of the turbulence and generate modes containing residual energy.  The expansion of the solar wind could also lead to the formation of magnetically dominated structures \citep{meyrand23} or influence the nonlinear interactions that generate residual energy \citep{shi23}.  While one or more of the aforementioned effects may play a role in residual energy generation, this paper will focus on isolating a single physical effect -- residual energy generated by overlapping, interacting Alfv{\'e}nic modes.  We will show that the reduced MHD equations that describe this system are symmetric in that they have no intrinsic preference for one sign of the residual energy over the other.  Initial and boundary conditions can break the symmetry; for the typically invoked initial value problem, this leads to net-negative residual energy generation.

We begin by deriving reduced MHD.  Our starting point is the incompressible MHD equations written in terms of Elsasser fields ${\bf z^{\pm}}=\delta{\bf v} \pm \delta{\bf b}$:

\begin{equation}
\frac{\partial{\bf z^{\pm}}}{\partial t} \mp \left ( {\bf V_A} \cdot \nabla \right ) {\bf z^{\pm}} = - {\bf z^{\mp}} \cdot \nabla {\bf z^{\pm}} - \frac{\nabla{P}}{\rho_0}
\label{eqn:inMHD}
\end{equation}

The pressure term in Eq.~\ref{eqn:inMHD} is determined by the incompressibility condition $\nabla \cdot {\bf z^\pm}=0$ together with the divergence of Eq.~\ref{eqn:inMHD} \citep{howes13}:

\begin{equation}
\frac{\nabla^2{P}}{\rho_0}=-\nabla \cdot \left( {\bf z^{\mp}} \cdot \nabla {\bf z^{\pm}} \right)
\end{equation}

Eq.~\ref{eqn:inMHD} describes not only Alfv{\'e}nic modes, but also slow modes and entropy fluctuations \citep{howes13}.  Because the latter two are found to be subdominant and passively advected in the inertial range of solar wind turbulence, a reduced MHD model which retains only the dominant Alfv{\'e}nic component is often employed \citep{schekochihin09, howes13, oughton17}.  In this model, which is rigorously valid in the anisotropic limit ($k_\perp \gg k_\parallel$, see Section II.C of \citet{howes13} for a full discussion), the component of magnetic and velocity fluctuations parallel to the mean magnetic field is set to zero.  We can then write Eq.~\ref{eqn:inMHD} in terms of Elsasser potentials $\zeta_\pm$, which are defined as \citep{schekochihin09, howes13}:

\begin{equation}
{\bf z^{\pm}} = {\bf \hat{z}} \times \nabla_\perp\zeta_\pm \label{eqn:zetadef}
\end{equation}

\noindent where ${\bf \hat{z}}$ represents the unit vector along the mean magnetic field, defined as pointing to the right; this is not to be confused with ${\bf z^{\pm}}$ representing the Elsasser fields.  $\zeta_\pm$ can be related to ${\bf \delta{v}}$ and ${\bf \delta{b}}$ \citep{howes13}:

\begin{subequations}
\begin{eqnarray}
{\bf \delta{v}_\perp} = {\bf\hat{z}}\times\nabla_\perp\frac{1}{2}(\zeta_++\zeta_-) \label{eqn:ufromzeta}\\
{\bf \delta{b}_\perp} = {\bf\hat{z}}\times\nabla_\perp\frac{1}{2}(\zeta_+-\zeta_-) \label{eqn:bfromzeta}
\end{eqnarray}
\end{subequations}

We will also find it useful to define $\phi_\pm$ as a new parallel coordinate in the frame moving in the $\mp \bf{\hat{z}}$ direction  along the mean magnetic field at the Alfv{\'e}n speed \citep{howes13}:

\begin{equation}
\phi_\pm = z \pm V_A{t} \label{eqn:phidef}
\end{equation}

\noindent where $z$ is the coordinate along the mean magnetic field.

With these definitions in place, we can take the curl of Eq.~\ref{eqn:inMHD} and substitute in Eqs.~\ref{eqn:zetadef} and \ref{eqn:phidef} to obtain the Elsasser Potential Equation \citep{schekochihin09,howes13}:

\begin{equation}
\frac{\partial\nabla_\perp^2\zeta_\pm}{\partial \phi_\mp} = \pm\frac{1}{4V_A}\left[ \{ \zeta_{+}, \nabla_\perp^2\zeta_{-} \}+\{ \zeta_{-}, \nabla_\perp^2\zeta_{+} \}\mp\nabla_\perp^2\{ \zeta_{+}, \zeta_{-} \} \right]\label{eqn:epoteq}
\end{equation}

\noindent where the Poisson bracket is defined as:

\begin{equation}
\{f,g\} = {\bf\hat{z}} \cdot \left( \nabla_\perp{f} \times \nabla_\perp{g} \right)\label{eqn:pbracket}
\end{equation}

\noindent and the parallel coordinates are changed to the variables $\phi_\pm$ via the method of characteristics used in \citet{howes13}; with this change, only a single parallel coordinate appears in each version of Eq.~\ref{eqn:epoteq}.  Physically, this means that the nonlinear distortion of $\zeta_\pm$ due to interactions with $\zeta_\mp$ takes place in the moving frame where the parallel coordinate is $\phi_\mp$.  When $\zeta_\mp$ is zero, the terms on the right hand side of Eq.~\ref{eqn:epoteq} that lead to this nonlinear distortion vanish, and $\zeta_\pm$ can therefore no longer depend on $\phi_\mp$.  This linear solution for $\zeta_\pm$ may still depend on the other parallel coordinate $\phi_\pm$; such a solution describes normal mode Alfv{\'e}n waves propagating in the $\mp \bf{\hat{z}}$ direction along the mean magnetic field with the dispersion relation $\omega=\mp{k_\parallel}V_A$.

The construction of a more general solution to Eq.~\ref{eqn:epoteq} is not as trivial.  Many prior attempts make use of a weak turbulence (WT) approximation \citep{galtier00, galtier01, nazarenko11, schekochihin22} in which, unlike the solar wind, the timescale associated with the nonlinear terms ($\tau_{nl} \equiv (k_\perp |z^\mp|)^{-1}$) on the right side of Eq.~\ref{eqn:inMHD} is assumed to be long compared to the timescale associated with the linear terms ($\tau_A \equiv (k_\parallel V_A)^{-1}$) on the left side.  Such models often assume that only normal modes that satisfy the Alfv{\'e}n wave dispersion relation are present; this condition, together with the frequency and wavenumber matching relations for three-wave interactions, implies that one of the three normal modes has $\omega=k_\parallel{V_A}=0$.  This mode does not oscillate in time and has no variation parallel to the mean magnetic field; it consists of stationary fluctuations that describe variation only in the directions perpendicular to the mean magnetic field.  We will therefore call this special normal mode a stationary 2D normal mode.  Thus WT describes interactions of finite $k_\parallel$ modes with $k_\parallel=0$ stationary 2D normal modes to produce secondary modes with the original $k_\parallel$, i.e.~energy is only transferred in the direction perpendicular to the mean magnetic field \citep{montgomery81,oughton94,ng96,galtier00,meyrand16}.  It has been shown in the context of WT theory that residual energy is spontaneously generated as a result of the nonlinear interaction \citep{boldyrev12a}, and that the negative residual energy condenses around a narrow region in phase space near $k_\parallel=0$ \citep{boldyrev09, wang11}.  However, the $k_\parallel=0$ modes that contain residual energy \citep{wang11} have an infinite linear timescale in this formalism, breaking the WT approximation, and reducing the utility of this approach \citep{schekochihin22}.

Residual energy has also been studied in the context of turbulence scaling theories, with the goal of deriving predictions for the residual energy spectrum.  One family of relevant models relies on an eddy-damped quasi-normal Markovian approximation in which residual energy is generated by the nonlinear interactions (\citet{grappin16} refers to this as a dynamo effect) and removed from the system on a timescale of $\tau_A$ by an Alfv{\'e}n effect (defined later in this paragraph) associated with the linear terms \citep{pouquet76,grappin82,grappin83}.  Subsequent refinements to the model incorporate the anistropy of MHD turbulence with respect to the mean magnetic field \citep{boldyrev12, gogoberidze12} and use $\tau_{nl}$ as the timescale for removal of residual energy \citep{grappin16}; the latter modification was made to better connect with critically balanced strong turbulence ($\tau_{nl} \sim \tau_A$), as found in the solar wind.  Predictions for the solar wind residual energy spectrum have also been made from numerical simulations \citep{boldyrev11, shi23}, and both theory and simulations show consistency with solar wind observations \citep{chen13}.  However, there is a potential problem with tying the residual energy relaxation rate to the linear Alfv{\'e}n wave time $\tau_A$.  According to this Alfv{\'e}n effect  formalism, first described by \citet{kraichnan65}, a newly formed nonlinear mode can be decomposed into forward and backward propagating Alfv{\'e}n normal mode waves.  The two waves decorrelate as they propagate away from each other, leading to a state in which the residual energy is asymptotically zero.  But this physical picture assumes the newly formed nonlinear mode is not supported by local plasma conditions, i.e. the mode is not in a region where a nonlinear drive term in the governing equation is present to sustain it.  This may not always be the case in the turbulent bath of fluctuations that make up the solar wind.

Our approach to the question of residual energy generation will instead rely on an examination of the reduced MHD equations inspired by the approach of \citet{howes13}.  In \citet{howes13}, Eq.~\ref{eqn:epoteq} is solved up to third order in $\tau_A/\tau_{nl}$ for the case of two overlapping Alfv{\'e}n waves in a periodic box and the result is benchmarked against numerical simulations \citep{nielson13}.  While this formalism is not a scaling theory and the authors do not analytically derive a turbulent spectrum, the results provide important insight into the underlying wave modes.  The second order solution of \citet{howes13} includes finite-frequency, purely magnetic modes with $k_\parallel=0$ that the authors attempt to tie to residual energy generation, but there are also finite-parallel-wavelength, $\omega=0$ modes present that are purely kinetic.  Their third order solution contains additional modes that are also not normal modes of the system -- despite having correlated magnetic and velocity fluctuations, these modes only exist in the presence of the nonlinear drive from the two initial Alfv{\'e}n waves and do not satisfy an Alfv{\'e}n wave dispersion relation.  We therefore term these nonresonant modes Alfv{\'e}nic quasimodes; analogous terminology has been previously used in the fusion community \citep{porkolab78, takase85}.  Modes that fit this description have also been seen in Alfv{\'e}n wave interaction experiments \citetext{\citealp{drake16}; C.~H.~K.~Chen, 2024} (where they are sometimes referred to as beat modes \citep{drake16}) as well as 3D MHD turbulence simulations in which a 4D Fourier transform was applied to analyze the spectral content of the various waves and structures \citep{yang19,markovskii20}.  Inclusion of quasimodes represents an important departure from WT theory, as we expect these nonlinear modes may play more of a role in the strong turbulence regime.  Note that the stationary 2D normal mode, which is also present in the solution of \citet{howes13}, is not a quasimode according to this definition because it satisfies $\omega=k_\parallel{V_A}=0$.

Prior work also suggests that residual energy may be intimately tied to initial and boundary conditions.  Simulations which find residual energy generation via nonlinear interactions are commonly conducted under periodic boundary conditions and initialize several overlapping wave modes, e.g.~\citep{oughton94, muller05, bigot08, boldyrev09, mininni09}.  Meanwhile, simulations by \citet{verniero18} consider both the periodic case of two initially overlapping Alfv{\'e}n waves and a case in which two Alfv{\'e}n wave packets initialized in distinct regions of space propagate towards each other.  While the first case generates multiple Alfv{\'e}nic quasimodes, after the wavepackets pass through each other in the second setup, all the resulting modes examined are Alfv{\'e}n normal modes with zero residual energy.  Although this is consistent with the fact that once the nonlinear drive is no longer present the plasma can only support normal mode waves, a physically intuitive explanation for the connection between initial conditions and residual energy generation remains elusive.

This paper aims to provide such a connection.  We show in Section~\ref{sec:esym} that the reduced MHD equations have no preference between net-positive and net-negative residual energy due to a symmetry that has not yet been reported in the literature.  In Section~\ref{sec:ival}, we show that when initial conditions which are not an exact solution at $t=0$ are applied, this important symmetry is broken.  A Fourier decomposition of Eq.~\ref{eqn:epoteq} then has both a particular solution at the frequency of the nonlinear drive and a homogeneous solution at the frequency of the associated normal mode. At the resonance where the two frequencies match, secularly growing Alfv{\'e}n normal modes are produced.  In Section~\ref{sec:renergy}, we show that both the homogeneous and secular solutions are highly dependent on the initial conditions in a way that breaks the symmetry of the governing equations, leading to net-negative residual energy generation by scale-local interactions.  For the chosen initial condition, this net-negative residual energy manifests in two ways: i) stationary 2D normal modes in the homogeneous and secular solutions are purely magnetic and ii) secular modes grow in time (not in space), leading to a time-dependent nonlinear drive that preferentially produces Alfv{\'e}nic quasimodes with negative residual energy.  The equivalent boundary value problem with a condition specified at $z=0$ generates net-positive residual energy; thus our results suggest the solar wind, where net-negative residual energy is observed \citep{matthaeus82, bavassano98, chen13}, is better described by an initial value problem in which a turbulent plasma parcel evolves nonlinearly from its initial state at $t=0$.  Important definitions used in the paper are summarized in Tables~\ref{tab:modedefs} and \ref{tab:variables} and the conclusions are discussed in Section~\ref{sec:concl}.

\begin{table}
\caption{\label{tab:modedefs}Table summarizing the definitions of key modes, interaction types, and transformations discussed in the text.}
\setlength{\tabcolsep}{2mm}
\begin{tabular}{p{4cm}|p{11.5cm}}
Term & Definition \\
\hline
\rowcolor{gray!40}
Alfv{\'e}n normal mode & Alfv{\'e}n wave solution to Eq.~\ref{eqn:inMHD} that satisfies the linear dispersion relation \\
Alfv{\'e}nic quasimode & Nonlinearly driven, non-normal mode that need not satisfy an Alfv{\'e}n wave dispersion relation but still retains some Alfv{\'e}nic properties such as incompressibility and a high degree of correlation between ${\bf \delta{b}}$ and ${\bf \delta{v}}$ \\
\rowcolor{gray!40}
Stationary 2D normal mode & Special Alfv{\'e}n normal mode with $\omega=k_{\parallel}V_A=0$ \\
Resonant interaction & Interaction in which the secondary mode produced is an Alfv{\'e}n normal mode  \\
\rowcolor{gray!40}
Nonresonant interaction & Interaction in which the secondary mode produced is an Alfv{\'e}nic quasimode \\
Resonant triad interaction & Interaction in which all three modes are Alfv{\'e}n normal modes (and one is a stationary 2D normal mode) \\
\rowcolor{gray!40}
Scale-local interaction & Interaction in which the two primary modes have similar perpendicular scales \\
Elsasser symmetric system & A system of interacting Alfv{\'e}nic modes that is invariant under at least one of two simultaneous variables interchanges i) ${\bf \delta{v}} \rightleftarrows {\bf \delta{b}}$, $z \rightleftarrows V_A{t}$, and an optional supplementary transformation and/or ii) ${\bf \delta{v}} \rightleftarrows -{\bf \delta{b}}$, $z \rightleftarrows -V_A{t}$, and an optional supplementary transformation.  Any linear superposition of non-interacting, Elsasser symmetric systems is also defined to be Elsasser symmetric. An Elsasser symmetric system has zero net residual energy. \\
\rowcolor{gray!40}
Supplementary transformation & Any coordinate translation, rotation, or reflection i) under which the reduced MHD equations are invariant, and ii) when applied the 4D Fourier transform of a system of interacting modes, only affects the sinusoidal phases of the modes 
\end{tabular}
\end{table}

\begin{table}
\caption{\label{tab:variables}Table of selected variables and operators used in this paper.  Listed are the equation or section that defines or introduces each variable or operator and a description of the quantity.}
\begin{ruledtabular}
\begin{tabular}{lll}
Variable or Operator & Equation & Description \\
\hline
${\bf\hat{z}}$ & Eq.~\ref{eqn:zetadef} & Unit vector along the mean magnetic field\\
$\bf z^{\pm}$ & $\delta{\bf v} \pm \delta{\bf b}$ & Elsasser fields\\
$\zeta_\pm$ & Eq.~\ref{eqn:zetadef} & Elsasser potentials\\
$\phi_\pm$ & Eq.~\ref{eqn:phidef} & Parallel coordinate in the $\omega=\mp k_\parallel V_A$ normal mode frame\\
$k_{\pm}$ & Eq.~\ref{eqn:kparpm} & Parallel wavenumber in the $\omega=\mp k_\parallel V_A$ normal mode frame\\
$E_v$ & Section~\ref{sec:esym} & Energy in velocity fluctuations \\
$E_b$ & Section~\ref{sec:esym} & Energy in magnetic fluctuations \\
$\rightarrow$ & Section~\ref{sec:esym} & Variable to the left of the arrow becomes the one on the right\\
$\rightleftarrows$ & Section~\ref{sec:esym} & Interchange variables on either side of double arrows\\
$\mathcal{N}_\pm$ & Eq.~\ref{eqn:epoteqop} & Reduced MHD nonlinear operator \\
$\mathcal{F}$ & Eq.~\ref{eqn:op_f} & Nonlinear Faraday operator \\
$\mathcal{M}$ & Eq.~\ref{eqn:op_m} & Nonlinear Momentum operator \\
$\mathcal{P}_\pm$ & Eq.~\ref{eqn:op_p} & Particular solution nonlinear operator \\
$\theta$ & Section~\ref{sec:esym} & Phase constant of a given mode \\
$\bar{\zeta}$ & Eq.~\ref{eqn:ft_op_n} & Fourier amplitude of $\zeta$ after transform over $\phi_\mp$ only \\
$\breve{\zeta}$ & Eq.~\ref{eqn:ivalsolc} & Fourier amplitude of $\zeta$ after transform over both $z$ and $t$ \\
$\zeta_{0\pm}$ & Eq.~\ref{eqn:ivalsetup} & Elsasser potentials at $t=0$\\
$\delta_k$ & Eq.~\ref{eqn:ivalsol} & Kronecker delta, $\delta_k=1$ if $k=0$, $\delta_k=0$ otherwise \\ 
$\delta(k)$ & Eq.~\ref{eqn:zeta_ab_onemode_k} & Dirac delta function\\
$\circledast$ & Eqs.~\ref{eqn:k_op_f}, \ref{eqn:k_op_m} & Convolution operator \\
$p_\pm(\ell,\omega,k_\parallel)$ & Eq.~\ref{eqn:pfn} & Parallel response function \\
$f_c(t)$ & Eq.~\ref{eqn:zeta_ab_onemode} & Time-dependent nonlinear drive amplitude \\
$f_c^{(n)}(t)$ & Eq.~\ref{eqn:ftdeltaomega} & $n$th derivative of $f_c(t)$\\
$\Delta\omega$ & Eq.~\ref{eqn:psolintc} & Any frequency for which $\breve{f}_c(\omega)$ is nonzero \\
$\theta_c$ & Eq.~\ref{eqn:zeta_ab_onemode} & Phase factor in the nonlinear drive\\
\end{tabular}
\end{ruledtabular}
\end{table}

\section{Elsasser Symmetry}\label{sec:esym}

The reduced MHD model described by Eq.~\ref{eqn:epoteq} has an important property that we will refer to as ``Elsasser symmetry:'' the equation is invariant under the simultaneous interchange of both i) magnetic and velocity fluctuations and ii) the coordinate parallel to the mean magnetic field with Alfv{\'e}n speed times the time coordinate.  In terms of variables that appear in Eq.~\ref{eqn:epoteq}, this may be written as a simultaneous negation of $\zeta_- \rightarrow -\zeta_-$ and $\phi_- \rightarrow -\phi_-$, per Eqs.~\ref{eqn:ufromzeta}, \ref{eqn:bfromzeta}, and \ref{eqn:phidef}.  Per the explanation that follows Eq.~\ref{eqn:pbracket}, the latter transformation corresponds to a reversal of the parallel coordinate in the frame where $\zeta_-$ distorts $\zeta_+$.  To keep the evolution of $\zeta_+$ unchanged, the sign of the nonlinear terms is also flipped via the negation of $\zeta_-$.  Meanwhile, the version of Eq.~\ref{eqn:epoteq} describing the distortion of $\zeta_-$ is also unchanged; there is no change to the parallel coordinate in the frame of the distortion, and all terms in Eq.~\ref{eqn:epoteq} pick up a cancellable negative sign.  This same argument for Elsasser symmetry applies to the other Elsasser potential: Eq.~\ref{eqn:epoteq} is also invariant under the simultaneous negation of variables $\zeta_+ \rightarrow -\zeta_+$ and $\phi_+ \rightarrow -\phi_+$.  Note that this symmetry is not present in the more general incompressible MHD formalism (Eq.~\ref{eqn:inMHD}) due to the retention of parallel gradients (and hence factors of $\phi_\pm$) in the nonlinear terms.  Reduced MHD equations that include viscosity and resistivity (e.g.~\citealp{oughton17}) will also not be Elsasser symmetric.

Importantly, the Elsasser symmetry of reduced MHD does not mean that every system described by Eq.~\ref{eqn:epoteq} will be Elsasser symmetric.  We define a reduced MHD system as Elsasser symmetric when the $\zeta_\pm$ functions describing the system are invariant under at least one of two simultaneous variable negations: i) $\zeta_- \rightarrow -\zeta_-$, $\phi_- \rightarrow -\phi_-$, and an optional supplementary transformation to be described later in this section and/or ii) $\zeta_+ \rightarrow -\zeta_+$, $\phi_+ \rightarrow -\phi_+$, and an optional supplementary transformation.  For simplicity, we will primarily consider simultaneous variable negation (i) in our calculations later in this paper, but our physical arguments will also be applicable to set (ii).

To more easily evaluate the effect of these negations, we will find it useful to conceptualize our system as a superposition of many interacting Fourier modes.  This may be achieved by applying a 4D Fourier transform over all three spatial dimensions and time; each mode then has a sinusoidal phase of ${\bf k_\perp} \cdot {\bf x_\perp} + k_\parallel{z}-\omega{t}+\theta$, where $\theta$ is a phase constant.  Per Eq.~\ref{eqn:phidef}, this sinusoidal phase will depend on $\phi_+$ and $\phi_-$; the $\phi_- \rightarrow -\phi_-$ (or $\phi_+ \rightarrow -\phi_+$) negation therefore acts on the sinusoidal phase of the modes.  Meanwhile, the $\zeta_- \rightarrow -\zeta_-$ (or $\zeta_+ \rightarrow -\zeta_+$) negation reverses the sign of mode amplitudes.  Here, we adopt the sign convention of \citet{howes13} in which the sign of ${\bf k}$ represents the mode propagation direction and $\omega$ is always positive.

Elsasser symmetry has important implications for the residual energy.  We define $E_v \sim (\delta{v})^2$ and $E_b \sim (\delta{b})^2$ as the energy in the velocity and magnetic fluctuations respectively.  The residual energy is then given by $E_r=E_v-E_b$.  The simultaneous negation of variables $\zeta_- \rightarrow -\zeta_-$ and $\phi_- \rightarrow -\phi_-$ may be equivalently written in a more intuitive form as the simultaneous interchange ${\bf \delta{v}} \rightleftarrows {\bf \delta{b}}$ and $z \rightleftarrows V_A{t}$, where $\rightleftarrows$ indicates that the symbols on either side of the operator are interchanged with each other.  Similarly, $\zeta_+ \rightarrow -\zeta_+$ and $\phi_+ \rightarrow -\phi_+$ may be expressed as ${\bf \delta{v}} \rightleftarrows -{\bf \delta{b}}$ and $z \rightleftarrows -V_A{t}$.  We note that under either simultaneous interchange of variables, not only does the form of Eq.~\ref{eqn:epoteq} remain the same, but so do the root mean square energy $<E_v+E_b>$ and cross helicity $<{\bf \delta{v}} \cdot {\bf \delta{b}}>$ for any constant-amplitude sinusoidal mode.  The mode's residual energy $<E_r>=<E_v-E_b>$, however, picks up a negative sign.  Therefore, in an Elsasser symmetric system of many such interacting Alfv{\'e}nic modes, individual modes may contain either positive or negative residual energy, but to be considered symmetric under the variable interchange, the system must have zero net residual energy.  Furthermore, because initial and boundary conditions in $t$ and $z$ will swap as a consequence of the variable interchange, they must be chosen to enable preservation of symmetry; for example, the system can be setup to be periodic in both $t$ and $z$ by choosing a set of constant-amplitude sinusoidal modes with phase arguments that depend linearly on both variables.  For solutions to Eq.~\ref{eqn:epoteq} that are not Elsasser symmetric, there is no reason based on the equation alone to prefer one sign of net residual energy over the other.  We will see in Section~\ref{sec:renergy} that the choice of initial conditions can break Elsasser symmetry in a way that favors net-negative residual energy generation.

Eq.~\ref{eqn:epoteq} has other symmetries that may need to be considered when evaluating whether a system is Elsasser symmetric.  For example, a translation of our coordinate system (e.g.~$\phi_+ \rightarrow \phi_\pm + \phi_{\pm0}$ or ${\bf x_\perp} \rightarrow {\bf x_\perp} + {\bf x_{\perp 0}}$, where $\phi_{\pm0}$ and ${\bf x_{\perp 0}}$ are constants) can change the mode phase constants; this has no effect on the physics of the system, as Eq.~\ref{eqn:epoteq} depends only on derivatives and is therefore translation invariant.  Similarly, Eq.~\ref{eqn:epoteq} is also invariant under a rotation of the perpendicular coordinate axes or under a reflection with respect to a line in the perpendicular plane.  The latter yields no net change because it negates both the cross product and one perpendicular gradient in the Poisson bracket.  We define any such coordinate translation, rotation, or reflection under which Eq.~\ref{eqn:epoteq} is invariant and which only affects the sinusoidal phase of the modes as a ``supplementary transformation.''  For some systems, including the example in the last paragraph of this section, the simultaneous interchange of variables ${\bf \delta{v}} \rightleftarrows {\bf \delta{b}}$ and $z \rightleftarrows V_A{t}$ (or ${\bf \delta{v}} \rightleftarrows -{\bf \delta{b}}$ and $z \rightleftarrows -V_A{t}$) will only produce the same system up to the sinusoidal phase of the modes.  If we can recover the original system by also adding in a supplementary transformation, then the system is still considered to be Elsasser symmetric.

We also define a system as Elsasser symmetric if the system can be expressed as a linear superposition of multiple non-interacting systems, and each individual system in the superposition is by itself Elsasser symmetric.  An example of this is a system of modes with aligned perpendicular wavenumbers, for which the nonlinear terms in Eq.~\ref{eqn:epoteq} are identically zero.  Each mode in the system then satisfies Eq.~\ref{eqn:epoteq} independently, the phases of these independent modes need not be related, and  there are no frequency and wavenumber matching conditions to satisfy.  As we will see in an example at the end of this section, this may make it impossible to design a supplementary transformation that can be simultaneously applied to every mode.  But because the equations for these modes are decoupled, it makes physical sense to treat each non-interacting mode as its own separate system for the purposes of evaluating Elsasser symmetry.

When interacting modes are present, the right side of Eq.~\ref{eqn:epoteq} represents a nonlinear drive due to Alfv{\'e}nic mode coupling and the left side represents the Alfv{\'e}nic modes that appear in the plasma in response.  We can therefore write Eq.~\ref{eqn:epoteq} in terms of a nonlinear operator $\mathcal{N}$:

\begin{subequations}
\begin{eqnarray}
\frac{\partial\zeta_\pm}{\partial \phi_\mp} = \mathcal{N_\pm}(\zeta_+,\zeta_-) &=& \mathcal{F}(\zeta_+,\zeta_-) \pm \mathcal{M}(\zeta_+,\zeta_-) \label{eqn:epoteqop}\\
\mathcal{F}(\zeta_+,\zeta_-) &=& \frac{-1}{4V_A}\{ \zeta_{+}, \zeta_{-} \} \label{eqn:op_f}\\
\mathcal{M}(\zeta_+,\zeta_-) &=& \frac{(\nabla_\perp^2)^{-1}}{4V_A}\left[ \{ \zeta_{+}, \nabla_\perp^2\zeta_{-} \}+\{ \zeta_{-}, \nabla_\perp^2\zeta_{+} \} \right] \label{eqn:op_m}
\end{eqnarray}
\end{subequations}

\noindent Here, $\mathcal{F}$ contains the nonlinear terms that arise from the curl of $v \times B$ in Faraday's law while $\mathcal{M}$ contains terms from the nonlinear convective derivative and nonlinear $j \times B$ term in the ion momentum equation.  This may be seen by noting that i) the ion momentum equation can be recovered by adding together the ${\bf z^+}$ and ${\bf z^-}$ versions of Eq.~\ref{eqn:inMHD} while Faraday's Law can be recovered from a difference between the two and ii) Due to the change of parallel coordinate variables, Eq.~\ref{eqn:epoteq} is multiplied by an extra $\mp$ sign compared to Eq.~\ref{eqn:inMHD}.

To build our intuition for Elsasser symmetric systems, it is useful to consider the case in which $\zeta^+$ and $\zeta^-$ are of similar perpendicular scales.  The two nonlinear terms in the $\mathcal{M}$ operator then approximately cancel each other, and the remaining $\mathcal{F}$ operator is identical for both sign choices in Eq.~\ref{eqn:epoteqop}.  This means that when modes of similar perpendicular scale interact, the ${\partial\zeta_+}/{\partial \phi_-}$ response and the ${\partial\zeta_-}/{\partial \phi_+}$ response are approximately the same.  Considering only the response at a single frequency and parallel wavenumber, we can use this result together with Eqs.~\ref{eqn:ufromzeta}, \ref{eqn:bfromzeta}, and \ref{eqn:phidef} to directly relate the phase speed of the nonlinearly driven mode to $\delta{v}/\delta{b}$:

\begin{equation}
r_A = \left( \frac{\delta{v}}{\delta{b}} \right )^2 \approx \left(\frac{k_\parallel V_A}{\omega} \right ) ^2 \label{eqn:rA}
\end{equation}

\noindent The quantity $r_A$ is defined as the Alfv{\'e}n ratio\footnote{This result was derived in the context of an interaction between modes of similar perpendicular scale in nonlinear reduced MHD, but Eq.~\ref{eqn:rA} exactly holds for a linear Alfv{\'e}nic fluctuation in the two-fluid regime.  It can be derived in the latter context by equating i) the cross-field ion current in an Alfv{\'e}nic mode with the ion polarization drift [Using the ion momentum equation (Eq.~30 of \citet{hollweg99}) and Amp{\`e}re's Law] and ii) the velocity fluctuations with the ion $E \times B$ drift.}.

For an MHD Alfv{\'e}n wave with $\omega=\mp k_\parallel V_A$, the residual energy is zero, and Eq.~\ref{eqn:rA} gives an Alfv{\'e}n ratio of $1$.  However, Alfv{\'e}nic quasimodes need not satisfy $\omega=\mp k_\parallel V_A$.  When modes of similar perpendicular scale interact, Eq.~\ref{eqn:rA} predicts that secondary modes with phase speeds less that the Alfv{\'e}n speed (i.e.~$|\omega|<|k_\parallel{V_A}|$) contain excess kinetic energy and modes with phase speeds greater than the Alfv{\'e}n speed (i.e.~$|\omega|>|k_\parallel{V_A}|$) contain excess magnetic energy.  We will see in Section~\ref{sec:renergy} how this generalizes to the interaction of modes that may not have similar perpendicular scales.  Under the interchange of variables ${\bf \delta{v}} \rightleftarrows {\bf \delta{b}}$ and $z \rightleftarrows V_A{t}$ (or ${\bf \delta{v}} \rightleftarrows -{\bf \delta{b}}$ and $z \rightleftarrows -V_A{t}$), a mode with positive residual energy that predominately varies in the parallel direction transforms into a mode with negative residual energy that predominately varies in time (and vice versa).  A system with no net residual energy contains both kinds of quasimodes in equal proportion.  We therefore expect such a system will be Elsasser symmetric, but as will be discussed in Section~\ref{sec:concl}, proof of this conjecture is left to future work.  This system may also contain $\omega=\mp k_\parallel V_A$ modes which retain the same dispersion relation under the variable interchange and therefore can have no residual energy.  This property applies to stationary 2D normal modes, suggesting that previously studied systems which include a 2D condensed region with large amounts of negative residual energy (e.g.~\citealp{wang11}) are not Elsasser symmetric.

To further demonstrate how to evaluate the requirements for Elsasser symmetry, consider one final example: a system that includes normal mode ``p'' which appears only on $\zeta_+$ and normal mode ``m'' which appears only on $\zeta_-$.  Mode ``p'' [``m''] is proportional to $\sin({\bf k_{p\perp}} \cdot {\bf x_\perp} - k_{p\parallel}\phi_+$) [$\sin({\bf k_{m\perp}} \cdot {\bf x_\perp} + k_{m\parallel}\phi_-)$].  We can see from these expressions that mode ``p'' [``m''] will satisfy the dispersion relation $\omega=-k_\parallel V_A$ [$\omega=k_\parallel V_A$].  Under the $\phi_- \rightarrow -\phi_-$ negation, the $\zeta_+$ normal mode ``p'' is unchanged, but the $\zeta_-$ mode ``m'' is now proportional to $-\sin(-{\bf k_{m\perp}} \cdot {\bf x_\perp} + k_{m\parallel}\phi_-)$.  To recover the same modes after applying the $\zeta_- \rightarrow -\zeta_-$ part of the simultaneous negation, a supplementary transformation involving the negation of an ${\bf x_\perp}$ component is clearly required, but this transformation must be designed to affect mode ``m'' but not mode ``p.''  It is possible to either satisfy or work around this requirement in three cases i) ${\bf k_{p\perp}} \parallel {\bf k_{m\perp}}$ and all modes in the system are non-interacting, ii) ${\bf k_{p\perp}} \perp {\bf k_{m\perp}}$, or iii) The direction of ${\bf k_{m\perp}}$ is not correlated with $k_{m\parallel}$.  In case (i) each individual mode can be considered as a separate system for the purpose of evaluating Elsasser symmetry and any supplementary transformation of mode ``m'' will therefore not affect mode ``p.''  Meanwhile, in case (ii) it is possible to design a supplementary transformation that affects mode ``m'' but not mode ``p'' by negating only the component of ${\bf x_\perp}$ parallel to ${\bf k_{m\perp}}$.  However, case (iii) is likely the most common in real systems as it does not require the angle between ${\bf k_{p\perp}}$ and ${\bf k_{m\perp}}$ to be an exact multiple of $\pi/2$.  In this case, even though mode ``m'' transforms into a mode with the opposite sign of ${\bf k_{m\perp}}$, there is an equal-amplitude mode in the system with this opposite sign of ${\bf k_{m\perp}}$ that transforms into mode ``m,'' preserving Elsasser symmetry.  This could occur, for example, in a system with no preferred ${\bf k_\perp}$ direction.  In case (ii) and likely also in case (iii), the full system will also include a large number of other modes that result from interactions involving modes ``p'' and ``m,'' and these modes must also be considered to determine the overall Elsasser symmetry of the system.

\section{The Initial Value Problem}\label{sec:ival}

In the remainder of this paper, we will examine how initial conditions can break Elsasser symmetry in a way that leads to net-negative residual energy.  We begin by considering Eq.~\ref{eqn:epoteq} in the context of an initial value problem where $\zeta_\pm=\zeta_{0\pm}$ at t=0.  Here, $\zeta_{0\pm}$ represents a set of constant-amplitude, sinusoidal Alfv{\'e}n normal modes polarized in various directions in the plane perpendicular to the mean magnetic field such that $\nabla_\perp\zeta_{0+} \times \nabla_\perp\zeta_{0-} \neq 0$.  This initial condition is chosen to represent typical initial simulation setups e.g.~\citep{oughton94, muller05, bigot08, boldyrev09, mininni09} in which modes with equal amounts of energy in magnetic and velocity fluctuations are initialized in a periodic box and net residual energy can subsequently develop.  Note that our initial condition is not an exact solution to Eq.~\ref{eqn:epoteq}, as secondary modes that result from the interaction between $\zeta_{0+}$ and $\zeta_{0-}$ are not included in the initial state.  By contrast, two non-overlapping wave packets traveling towards each other is an exact solution at $t=0$, as there is then no initial nonlinear drive; while we will briefly comment in this section on how the physics changes when the initial state is an exact solution to Eq.~\ref{eqn:epoteq}, a full treatment of this alternate case is left to future work.  With our chosen initial condition included, Eq.~\ref{eqn:epoteq} can be written as:

\begin{equation}
\zeta_\pm=\zeta_{0\pm} + \int_{\phi_\pm}^{\phi_\mp} \mathcal{N}_\pm(\zeta_+,\zeta_-)  d\phi'_\mp \label{eqn:ivalsetup}
\end{equation}

\noindent Integration of the $+$ mode takes place in the frame of the $-$ mode and vice versa.  This integration over $\phi'_\mp$ is from $\phi'_\mp=\phi_\pm$ (which corresponds to $t=0$) to an arbitrary time.  The $t=0$ limit of integration ensures that our initial condition consists of only $\zeta_{0\pm}$.  If $\zeta_{0\pm}$ is already an exact solution at $t=0$, this lower limit evaluates to zero.

The simple act of applying an initial condition to Eq.~\ref{eqn:epoteq} that is not an exact solution breaks Elsasser symmetry.  This important result can most easily be seen by considering the $\zeta_-$ version of Eq.~\ref{eqn:ivalsetup} under the simultaneous negation $\zeta_- \rightarrow -\zeta_-$ and $\phi_- \rightarrow -\phi_-$.  As explained at the start of Section~\ref{sec:esym}, this simultaneous negation simply adds a cancellable negative sign to both sides of Eq.~\ref{eqn:epoteq}; but in Eq.~\ref{eqn:ivalsetup}, $\zeta_{0-}$ does not pick up this negative sign.  More importantly, the lower limit of integration at $\phi'_+=\phi_-$ corresponding to $t=0$ becomes $\phi'_+=-\phi_-$, which corresponds to $z=0$.  Thus, our initial value problem has transformed into a boundary value problem with $-\zeta_{0-}$ as the boundary condition on $\zeta_-$ at $z=0$.  This broken Elsaaser symmetry implies that terms arising from the asymmetric lower limit of integration imposed to satisfy our initial condition will play a key role in residual energy generation.

To gain additional insight into the form and role of these terms, we will find it useful to Fourier transform the nonlinear operator with respect to the normal mode frame coordinate $\phi_\mp$; this transformation corresponds to a description of the system as a set of interacting Fourier modes.  We will employ a continuous Fourier transform in this paper, corresponding to a system that is infinitely large in both the parallel direction and time; this choice makes it as easy as possible to perform the interchange $z \rightleftarrows V_A{t}$ that is part of the test for Elsasser symmetry.  A periodic box considered for all time is a subset of this type of system, as the box can be made infinite in the parallel direction by infinite repetition of the periodic pattern.  We can then write $\mathcal{N}_\pm(\zeta_+,\zeta_-)$ as:

\begin{equation}
\mathcal{N}_\pm(\zeta_+,\zeta_-)=\frac{1}{\sqrt{2\pi}}\int_{-\infty}^\infty \mathcal{\bar{N}_\pm}(\zeta_+,\zeta_-) e^{ik_\mp\phi_\mp} dk_\mp \label{eqn:ft_op_n}
\end{equation}

\noindent Here, $\mathcal{\bar{N}_\pm}(\zeta_+,\zeta_-)$ is a function of $k_\mp$ representing the Fourier transform of $\mathcal{N_\pm}(\zeta_+,\zeta_-)$; this Fourier amplitude may depend on the perpendicular coordinates and on $\phi_\pm$, but not on $\phi_\mp$.   The wavenumber in this coordinate system $k_\pm$ is defined by equating the sinusoidal phase $k_\parallel{z}-\omega{t}$ with $k_+\phi_+ + k_-\phi_-$:

\begin{equation}
k_\pm = \frac{1}{2} \left( k_\parallel \mp \frac{\omega}{V_A} \right) \label{eqn:kparpm}
\end{equation}

Plugging Eq.~\ref{eqn:ft_op_n} with $\phi_\mp=\phi'_\mp$ into Eq.~\ref{eqn:ivalsetup} and performing the integration over $\phi'_\mp$:

\begin{equation}
\begin{aligned}
\zeta_\pm=&-\frac{i}{\sqrt{2\pi}}\int_{-\infty}^\infty \left[1-\delta_{k_\mp}\right] \frac{\mathcal{\bar{N}_\pm}(\zeta_+,\zeta_-)}{k_\mp} e^{ik_\mp\phi_\mp} dk_\mp \\
&+ \zeta_{0\pm} + \frac{i}{\sqrt{2\pi}}\int_{-\infty}^\infty \left[1-\delta_{k_\mp}\right] \frac{\mathcal{\bar{N}_\pm}(\zeta_+,\zeta_-)}{k_\mp} e^{ik_\mp\phi_\pm} dk_\mp \\
&\mp (\phi_+ - \phi_-)  \mathcal{\bar{N}_\pm}(\zeta_+,\zeta_-)\bigg|_{k_\mp=0} \label{eqn:ivalsol}
\end{aligned}
\end{equation}

We identify three distinct parts of the solution in Eq.~\ref{eqn:ivalsol}.  The term on the first line is from the upper limit of integration in Eq.~\ref{eqn:ivalsetup} and retains the same $k_\mp\phi_\mp$ dependence as the nonlinear drive in Eqs.~\ref{eqn:ivalsetup} and \ref{eqn:ft_op_n}.  This term therefore represents the response of the plasma at the frequency and wavenumber of the nonlinear drive, which is called the particular solution.  The resulting Alfv{\'e}nic quasimodes are not constrained by the Alfv{\'e}n dispersion relation.  By contrast, the terms on the second line have no $\phi_\mp$ dependence, but depend on $\phi_\pm$ both explicitly and through the $\mathcal{\bar{N}_\pm}(\zeta_+,\zeta_-)$ Fourier amplitude and $\zeta_{0\pm}$ initial condition.  These $\zeta_\pm$ modes are therefore normal modes satisfying the Alfv{\'e}n dispersion $\omega=\mp k_\parallel{V_A}$.  This response at the normal mode frequency, known as the homogeneous solution, includes both the normal modes in the initial condition and an integral term; the latter is present due to the lower limit of integration imposed to satisfy the initial condition at $t=0$.  Note that this second homogeneous term is not needed if the chosen initial conditions $\zeta_{0\pm}$ are already an exact solution to Eq.~\ref{eqn:epoteq}.  These definitions of the particular and homogeneous solutions are taken from the mathematics of differential equations \citep{edwards18}.  Finally, the third line combines terms from the upper and lower integration limits into a $\phi_+ - \phi_- = 2 V_A t$ dependence, which as we will see in Section~\ref{sec:homosec} is problematic for Elsasser symmetry.  This secularly growing term contains interactions between mode pairs with $k_\mp$ values that add up to zero, which corresponds to the case where the nonlinear drive is at the normal mode frequency and wavenumber.  A factor of one minus the Kronecker delta function $\delta_{k_\mp}$ is used in the particular and homogeneous solutions to exclude this resonance. 

\begin{figure}
\includegraphics[width=\textwidth]{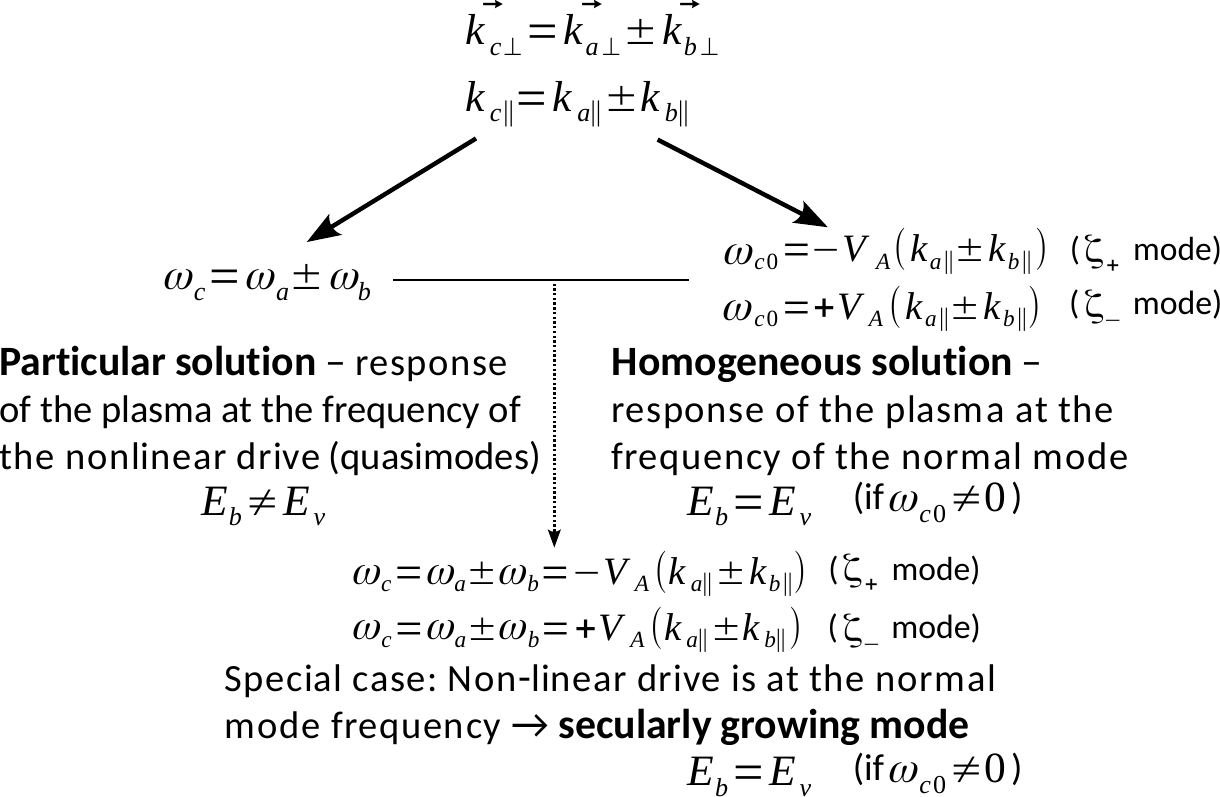}
\caption{\label{fig:soltypes} Three different components of the solution (particular, homogeneous, secular) described by Eq.~\ref{eqn:ivalsolc}.  Mode ``c'' represents the secondary mode that arises from the interactions of primary modes ``a'' and ``b.''  Since $\zeta_+$ ($\zeta_-$) normal modes propagate to the left (right) with the dispersion relation $\omega=-k_\parallel{V_A}$ ($\omega=+k_\parallel{V_A}$), each of these two cases is noted for the homogeneous and secular solutions.  Alfv{\'e}nic quasimodes that make up the particular solution need not follow the Alfv{\'e}n wave dispersion relation.  The special case of the stationary 2D normal mode ($\omega_{c0}=0$) is discussed in Section~\ref{sec:homosec}.}
\end{figure}

The fact that the solution has these three distinct parts is a key result of this paper, summarized in Fig.~\ref{fig:soltypes}.  For conceptual simplicity, Fig.~\ref{fig:soltypes} considers only a subset of the system described by Eq.~\ref{eqn:ivalsol}: a single Fourier mode ``a'' interacts with Fourier mode ``b'' to produce mode ``c.''  To better relate Eq.~\ref{eqn:ivalsol} to Fig.~\ref{fig:soltypes}, we perform an additional Fourier transform of Eq.~\ref{eqn:ivalsol} over $\phi_\pm$ and note that $\int_{-\infty}^\infty \int_{-\infty}^\infty dk_{\parallel}d\omega = \int_{-\infty}^\infty \int_{-\infty}^\infty 2V_A dk_+ dk_-$ to obtain:

\begin{equation}
\begin{aligned}
\zeta_{\pm}=&-\frac{iV_A}{\pi}\int_{-\infty}^\infty \int_{-\infty}^\infty \left[1-\delta_{k_{\parallel}\pm\omega/V_A}\right] \frac{\mathcal{\breve{N}_\pm}(\zeta_+,\zeta_-)}{k_{\parallel}V_A\pm\omega} e^{i(k_{\parallel}{z}-\omega{t})}dk_\parallel d\omega \\
&+  \zeta_{0\pm}+ \frac{iV_A}{\pi}\int_{-\infty}^\infty \int_{-\infty}^\infty \left[1-\delta_{k_{\parallel}\pm\omega/V_A}\right] \frac{\mathcal{\breve{N}_\pm}(\zeta_+,\zeta_-)}{k_{\parallel}V_A\pm\omega} e^{ik_{\parallel}(z \pm V_A t)} dk_\parallel d\omega \\
&\mp V_A{t}\sqrt{\frac{2}{\pi}} \int_{-\infty}^\infty \mathcal{\breve{N}_\pm}(\zeta_+,\zeta_-)\bigg|_{\omega=\mp k_\parallel V_A}e^{ik_{\parallel}(z \pm V_A t)} dk_\parallel \label{eqn:ivalsolc}
\end{aligned}
\end{equation}

\noindent As in Eq.~\ref{eqn:ivalsol}, the three lines of Eq.~\ref{eqn:ivalsolc} represent the particular, homogeneous, and secular solutions respectively.  The notation $\mathcal{\breve{N}_\pm}(\zeta_+,\zeta_-)$ is used to indicate the 2D Fourier transform of the nonlinear operator result over both $z$ and $t$; this quantity absorbs the factor of $2V_A$ from the change in the variables of integration.  Resonant interactions produce the normal modes described by the third line of Eq.~\ref{eqn:ivalsolc} that grow with time; meanwhile, quasimodes described by the first line result from nonresonant interactions, which means that they appear immediately whenever the associated nonlinear drive is present.  This does not invalidate our chosen initial condition because the homogeneous solution cancels the particular solution at $t=0$.  To see this, note that the particular and homogeneous integrals in Eq.~\ref{eqn:ivalsolc} are identical except for the time dependence and an overall sign flip.  The factor of $\omega$ in the denominator of both solutions is the frequency of the quasimode in the particular solution.  The sinusoidal phase of $k_{\parallel}\left(z \pm V_A{t}\right) = \left(k_+ + k_-\right)\phi_\pm$ in the homogeneous term has the same parallel wavenumber as the nonlinear drive but is at the associated normal mode frequency, as indicated in Fig.~\ref{fig:soltypes}.  As long as $k_\parallel \neq 0$, this sinusoidal dependence will be different for the $\zeta_+$ and $\zeta_-$ solution, leading to two distinct normal modes.

We note here that WT theory does not contain or consider all three parts of the solution summarized in Fig.~\ref{fig:soltypes}.  Traditional WT theory \citep{galtier00, galtier01} considers only the interaction of resonant triads, which consist of normal modes satisfying the Alfv{\'e}n dispersion relation; the theory does not contain Alfv{\'e}nic quasimodes.  Therefore, the only interaction from our solution allowed in WT theories is a subset of the secular solution in which one of the interacting modes has $\omega=k_\parallel{V_A}=0$.  Alfv{\'e}nic quasimodes interacting to produce a secularly growing normal mode is not allowed in WT theory, as this set of modes does not form a resonant triad.

Also unlike WT theory, individual secular modes are not set up a priori as a closed-form solution with slowly varying amplitudes; to see this result in our formalism, we must instead consider multiple individual interactions which yield an expansion for the mode amplitude in powers of $t$; this will be the subject of a future publication.  In the present paper, we are not trying to construct a closed-form solution to Eq.~\ref{eqn:epoteq}; such a solution is not required to show the relationship between broken Elsasser symmetry due to initial conditions and net-negative residual energy generation.  It is therefore not necessary to adopt a random phase or closure approximation or assume that the nonlinear interactions remain weak.  Eqs.~\ref{eqn:ivalsetup}, \ref{eqn:ivalsol}, and \ref{eqn:ivalsolc} are therefore valid for any reduced MHD system that is well-described by our initial conditions and Fourier representation.

\section{Physics of Residual Energy Generation}\label{sec:renergy}

Since Elsasser symmetry is broken when our initial condition is applied in Eq.~\ref{eqn:ivalsetup}, we expect the system to have net residual energy.  We now examine the various terms in Eq.~\ref{eqn:ivalsolc} to see where this residual energy lies.  Note that per Eqs.~\ref{eqn:ufromzeta} and \ref{eqn:bfromzeta}, this means that we need to examine the difference between $\delta{v_\perp} \sim \zeta_+ + \zeta_-$ and $\delta{b_\perp} \sim \zeta_+ - \zeta_-$.  It follows that if a mode only contributes to a single Elsasser potential ($\zeta_+$ or $\zeta_-$), it does not contain residual energy.  We will examine the particular solution in Section~\ref{sec:particular} and the homogeneous and secular solutions in Section~\ref{sec:homosec} to show that while the particular solution can be Elsasser symmetric per the requirements in Section~\ref{sec:esym}, initial conditions break the symmetry of the homogeneous and secular solutions.  For scale-local interactions this symmetry breaking preferentially produces negative residual energy in stationary 2D normal modes.  Finally, in Section~\ref{sec:threemode} and \ref{sec:partsym} we show how the presence of secular terms that grow in space rather than time can, upon subsequent interactions, break the symmetry of the particular solution, also leading to net-negative residual energy generation in scale-local interactions.

\subsection{Particular Solution}\label{sec:particular}

To examine the residual energy in the particular solution, we rewrite the first line of Eq.~\ref{eqn:ivalsolc} as the sum and difference of the two Elsasser potentials:

\begin{equation}
\begin{aligned}
\left(\zeta_{+} \pm \zeta_{-}\right)_{particular} = &\frac{V_A}{\pi}\int_{-\infty}^\infty \int_{-\infty}^\infty \left[1-\delta_{k_{\parallel}^2-(\omega/V_A)^2}\right] \mathcal{\breve{P}}_\pm(\zeta_+,\zeta_-) e^{i(k_{\parallel}{z}-\omega{t})}dk_\parallel d\omega \\
&-\frac{i}{\sqrt{2\pi}}\int_{-\infty}^\infty \left[1-\delta_{k_\parallel}\right] \frac{\mathcal{\breve{N}_+}(\zeta_+,\zeta_-)}{k_\parallel}\bigg|_{\omega= k_\parallel V_A}e^{ik_{\parallel}(z - V_A t)} dk_\parallel \\
&\mp\frac{i}{\sqrt{2\pi}}\int_{-\infty}^\infty \left[1-\delta_{k_\parallel}\right] \frac{\mathcal{\breve{N}_-}(\zeta_+,\zeta_-)}{k_\parallel}\bigg|_{\omega= -k_\parallel V_A}e^{ik_{\parallel}(z + V_A t)} dk_\parallel\label{eqn:ivalsolp}
\end{aligned}
\end{equation}

\noindent where the second and third terms represent $\zeta_+$ quasimodes moving to the right with $\omega=k_\parallel{V_A}$ and $\zeta_-$ quasimodes moving to the left with $\omega=-k_\parallel{V_A}$ respectively.  Quasimodes that fit this description were recently observed in a numerical simulation where they were referred to as anomalous fluctuations \citep{yang23}.  These are not normal modes because for our choice of sign convention, a normal mode Alfv{\'e}n wave traveling to the right (left) must be polarized as $\zeta_-$ ($\zeta_+$) to be an exact solution to Eq.~\ref{eqn:inMHD}, which means that the mode has anti-correlated (correlated) magnetic and velocity fluctuations.  The combination of the $\zeta_-$ ($\zeta_+$) quasimode on line 3 (2) of Eq.~\ref{eqn:ivalsolp} and a $\zeta_+$ ($\zeta_-$) normal mode from the corresponding homegenous or secular solution in Eq.~\ref{eqn:ivalsolc} can then contain residual energy.  Per the argument in the second to last paragraph of Section~\ref{sec:esym}, any such residual energy breaks Elsasser symmetry; however, this residual energy is unlikely to be very important.  If the system is dominated by resonant interactions, the growing secular component of the mode in question, which contains no residual energy, will be significantly larger than the other terms.  Meanwhile, in a system dominated by nonresonant interactions, there are likely to be many more possible quasimodes with $\omega \neq\mp k_\parallel{V_A}$ than modes that satisfy the dispersion relation.  Thus, the remainder of this section will consider residual energy arising from the particular solution alone, which will come from the first line in Eq.~\ref{eqn:ivalsolp}.

This first line encompasses terms that contribute to both $\zeta_+$ and $\zeta_-$ in the particular solution.  Here, the particular solution nonlinear operator $\mathcal{\breve{P}}_\pm(\zeta_+,\zeta_-)$ is defined as:

\begin{equation}
 \mathcal{\breve{P}}_\pm(\zeta_+,\zeta_-) = i\mathcal{\breve{F}}(\zeta_+,\zeta_-) \left[\frac{-1}{\omega+k_\parallel{V_A}} \pm \frac{1}{\omega-k_\parallel{V_A}}\right] + i\mathcal{\breve{M}}(\zeta_+,\zeta_-)\left[\frac{-1}{\omega+k_\parallel{V_A}} \mp \frac{1}{\omega-k_\parallel{V_A}}\right]\label{eqn:op_p}
\end{equation}

Eqs.~\ref{eqn:ivalsolp} and \ref{eqn:op_p} together with Eqs.~\ref{eqn:ufromzeta} and \ref{eqn:bfromzeta} imply that for $\omega \neq \mp k_\parallel{V_A}$, $\delta{\breve{v}_\perp} \sim \mathcal{\breve{P}}_+(\zeta_+,\zeta_-)$ and $\delta{\breve{b}_\perp} \sim \mathcal{\breve{P}}_-(\zeta_+,\zeta_-)$.  In other words, the expressions for the magnetic and velocity fluctuations differ only by the combinations of $k_\parallel$ and $\omega$ present in the $\breve{P}_\pm$ operator.  These differences imply that a given quasimode can carry either negative or positive residual energy, which is indicated in Fig.~\ref{fig:soltypes} by the note that $E_b \neq E_v$ for each individual mode that only appears in the particular solution.  Because the sinusoidal dependence will be the same for both versions of Eq.~\ref{eqn:ivalsolp}, we also expect the magnetic and velocity fluctuations of Alfv{\'e}nic quasimodes to be highly correlated.

We next examine the Elsasser symmetry of the particular solution for $\omega \neq \mp k_\parallel{V_A}$, i.e.~keeping only the first line of Eq.~\ref{eqn:ivalsolp}.  The $z \rightleftarrows V_A{t}$ part of the variable interchange has the same effect as interchanging integration variables $\omega \rightleftarrows -k_\parallel{V_A}$ in the sinusoidal exponent.  To ensure this causes the solutions for magnetic and velocity fluctuations to swap, i.e.~$\zeta_+ + \zeta_- \rightleftarrows \zeta_+ - \zeta_-$, we therefore require that $\mathcal{\breve{P}}_+(\zeta_+,\zeta_-) \rightleftarrows \mathcal{\breve{P}}_-(\zeta_+,\zeta_-)$ under the variable interchange $\omega \rightleftarrows -k_\parallel{V_A}$ plus any optional supplementary transformation.  Examining the form of Eq.~\ref{eqn:op_p}, we can see that this will be true as long as the $\mathcal{\breve{F}}(\zeta_+,\zeta_-)$ and $\mathcal{\breve{M}}(\zeta_+,\zeta_-)$ operators (or equivalently the $\mathcal{\breve{N}}_\pm(\zeta_+,\zeta_-)$ operators) pick up a negative sign under the variable interchange $\omega \rightleftarrows -k_\parallel{V_A}$ plus optional supplementary transformation.  This is equivalent to saying that the inverse transform $\mathcal{N}_\pm(\zeta_+,\zeta_-)$ picks up a negative sign under the variable interchange $z \rightleftarrows V_A{t}$ plus optional supplementary transformation; this matches our expectations from the analysis of the Elsasser symmetry of Eq.~\ref{eqn:epoteq} at the start of Section~\ref{sec:esym}.  Therefore, the particular solution given by the first line of Eq.~\ref{eqn:ivalsolp} can be Elsasser symmetric, and it has the same requirements for Elsasser symmetry as the original Eq.~\ref{eqn:epoteq}.  This symmetry can be broken by initial conditions, as we will see in Section~\ref{sec:partsym}.

\subsection{Homogeneous and Secular Solutions}\label{sec:homosec}

The homogeneous and secular terms in Eq.~\ref{eqn:ivalsolc} are made up of normal modes which, as long as $k_{\parallel} \neq 0$ in the sinusoidal dependence, will contribute only to $\zeta_+$ or only to $\zeta_-$ and by themselves contain no net residual energy.  This is indicated in Fig.~\ref{fig:soltypes} by the note that $E_b=E_v$ for each of these normal modes in the solution.  However, there is also a special case corresponding to a stationary 2D normal mode noted.  We will now examine this special case to show that scale-local interactions preferentially produce net-negative residual energy in these $\omega=k_\parallel{V_A}=0$ modes.

A stationary 2D normal mode arises in the homogeneous solution when a $\zeta_+$ mode interacts with a $\zeta_-$ mode of the same $|k_\parallel|$, producing a $k_\parallel=0$ quasimode in the particular solution with finite $\omega$, and a corresponding $\omega=k_\parallel{V_A}=0$ mode in the homogeneous solution on both Elsasser potentials.  This $\omega=k_\parallel{V_A}=0$ mode is the $k_{\parallel}=0$ (i.e.~$k_+=-k_-$) contribution to the integral on the second line of Eq.~\ref{eqn:ivalsolc}.  Per the argument in the second to last paragraph of Section~\ref{sec:esym}, the presence of the stationary 2D normal mode on both $\zeta_+$ and $\zeta_-$ breaks Elsasser symmetry; thus it is not surprising that this mode carries residual energy.  By using Eq.~\ref{eqn:ivalsolc} to calculate $\zeta_+ \pm \zeta_-$, we can show that the velocity fluctuations in this mode are proportional to an integral over $\mathcal{\breve{M}}(\zeta_+,\zeta_-)/\omega$ while the magnetic fluctuations are proportional to an integral over $\mathcal{\breve{F}}(\zeta_+,\zeta_-)/\omega$, where $\omega$ is the frequency of the quasimode that our stationary 2D normal mode must cancel at $t=0$, and both operators are evaluated at $k_\parallel=0$.  A similar dependence may be seen for stationary 2D normal modes in the secular solution on line three of Eq.~\ref{eqn:ivalsolc}, with velocity fluctuations proportional to $\mathcal{\breve{M}}(\zeta_+,\zeta_-)$ and magnetic fluctuations proportional to $\mathcal{\breve{F}}(\zeta_+,\zeta_-)$, where both operators are evaluated at $\omega=k_\parallel{V_A}=0$.  These secular $\omega=k_\parallel{V_A}=0$ modes are produced when the interacting $\zeta_+$ and $\zeta_-$ modes have frequencies and parallel wavenumbers that both sum to zero or both subtract to zero.  Thus for both the homogeneous and secular solutions, residual energy in stationary 2D normal modes crucially depends on the difference between the $\mathcal{\breve{F}}$ and $\mathcal{\breve{M}}$ operators.

To better understand the difference between these operators, it is useful to consider a simplified situation in which modes ``a'' and ``b,'' which  each have a well-defined perpendicular wavenumber, interact to produce mode ``c.''  In this case, the Fourier transforms of Eqs.~\ref{eqn:op_f} and \ref{eqn:op_m} may be written as:

\begin{subequations}
\begin{eqnarray}
\mathcal{\breve{F}}(\zeta_{a+},\zeta_{b-}) &=& \frac{-{\bf\hat{z}}}{4 V_A} \cdot \left(\bf{k_{a\perp}} \times \bf{k_{b\perp}}\right) \breve{\zeta}_{a+} \circledast \breve{\zeta}_{b-} \label{eqn:k_op_f}\\
\mathcal{\breve{M}}(\zeta_{a+},\zeta_{b-}) &=& \frac{-{\bf\hat{z}}}{4 V_A} \cdot \left(\bf{k_{a\perp}} \times \bf{k_{b\perp}}\right) \frac{k_{a\perp}^2 - k_{b\perp}^2}{k_{c\perp}^2} \breve{\zeta}_{a+} \circledast \breve{\zeta}_{b-}\label{eqn:k_op_m}
\end{eqnarray}
\end{subequations}

\noindent where $\circledast$ represents the convolution operator.  We note that these expressions are consistent with derivations of the interaction coefficient that are often the first step in WT theories (e.g.~\citet{schekochihin22}~Eq.~A4,~A5, \citet{nazarenko11}~Eq.~14.9,~14.10).

Eqs.~\ref{eqn:k_op_f} and \ref{eqn:k_op_m} show that the $\mathcal{\breve{F}}$ and $\mathcal{\breve{M}}$ operators only differ by the important fraction $(k_{a\perp}^2-k_{b\perp}^2)/{k_{c\perp}^2}$. This term can also be expressed as $({\bf k}_{a\perp}+{\bf k}_{b\perp})\cdot({\bf k}_{a\perp}-{\bf k}_{b\perp})/|{\bf k}_{c\perp}|^2$, a dot product between two possible secondary mode wavenumbers.  This term could in principle have any value, however, with two critical assumptions it can be argued that it is small for most interactions in the system: i) Energy in our system is transferred from large to small scales, implying that for most interactions $|{\bf k_{c\perp}}| = \max(|{\bf k}_{a\perp}+{\bf k}_{b\perp}| , |{\bf k}_{a\perp}-{\bf k}_{b\perp}|)$ and therefore $(k_{a\perp}^2-k_{b\perp}^2)/{k_{c\perp}^2} \leq 1$ and ii) Scale locality: modes of similar perpendicular scale are more likely to interact than modes of disparate scales, so $k_{a\perp} \sim k_{b\perp}$.  Both of these assumptions are commonly employed by leading MHD turbulence theories \citep{goldreich95, boldyrev05}, and a theoretical proof of scale locality for a suitable power law scaling was performed by \citet{aluie10}.  The shell-to-shell (scale-to-scale) energy transfer function is found in simulations to primarily involve modes of similar perpendicular scales \citep{alexakis07,debliquy05}, although some non-local interactions are also present \citep{meyrand16}.  With these assumptions in place, we can see the magnetic fluctuations ($\sim \mathcal{\breve{F}}$) of the stationary 2D normal mode will be much greater than the velocity fluctuations ($\sim \mathcal{\breve{M}}$) because the velocity fluctuations are multiplied by an extra factor of $(k_{a\perp}^2-k_{b\perp}^2)/{k_{c\perp}^2}$.  When the interactions in question are precisely scale-local, this factor will be zero and the stationary 2D normal modes in the homogeneous and secular solutions will be purely magnetic.

Purely magnetic stationary 2D normal modes in the homogeneous and secular solutions are a direct consequence of the breaking of Elsasser symmetry by an initial condition that is not an exact solution to Eq.~\ref{eqn:epoteq}.  As shown in Section~\ref{sec:ival}, these modes arise in whole or in part due to the lower limit of integration in Eq.~\ref{eqn:ivalsetup} that is key to the symmetry breaking.  To better understand how this choice of an initial condition is related to residual energy, it is useful to consider what happens if in lieu of an initial condition that is not an exact solution at $t=0$ we implement a boundary condition that is not an exact solution at $z=0$; this switch is accomplished by changing the lower limit of integration in Eq.~\ref{eqn:ivalsetup} from $\phi_\pm$ to $-\phi_\pm$.  The phase argument in the homogeneous solution in Eq.~\ref{eqn:ivalsolc} then becomes $\pm(k_+-k_-)\phi_\pm=\mp\omega(z/V_A \pm t)$, leading to a stationary 2D normal mode when $\omega=0$ (i.e.~$k_+=k_-$).  Following the same procedure used in the second paragraph of this section, velocity fluctuations in this mode are proportional to an integral over $\mathcal{\breve{F}}(\zeta_+,\zeta_-)/k_\parallel$ while the magnetic fluctuations are proportional to an integral over $\mathcal{\breve{M}}(\zeta_+,\zeta_-)/k_\parallel$, where both operators are now evaluated at $\omega=0$, and $k_\parallel$ is the wavenumber of the quasimode that cancels our stationary 2D normal mode at $z=0$.  Therefore, for scale local interactions, we will now have purely kinetic stationary 2D normal modes in the homogeneous solution.  In the secular solution, the $\mp(\phi_+ - \phi_-) = \mp{2}V_A{t}$ factor in Eqs.~\ref{eqn:ivalsol} and \ref{eqn:ivalsolc} will change to $(\phi_+ + \phi_-)=2z$.  The sign change in this factor means that the velocity fluctuations in the secular solution are now proportional to $\mathcal{\breve{F}}(\zeta_+,\zeta_-)$ and magnetic fluctuations are proportional to $\mathcal{\breve{M}}(\zeta_+,\zeta_-)$, where both operators are evaluated at $\omega=k_\parallel{V_A}=0$.  An $\omega=k_{\parallel} V_A = 0$ mode, which grows in space rather than in time, will now be purely kinetic when it is produced by scale local interactions.

Secular modes with $\omega=\mp k_{\parallel} V_A \neq 0$ also pose a problem for Elsasser symmetry.  Secular modes grow in time, which means that the system must also include corresponding modes that grow in space in order to be Elsasser symmetric.  But, as is evident from Eq.~\ref{eqn:ivalsolc}, in an initial value problem with a  nonlinear drive at $t=0$, only modes that vary with time are produced at resonance.  Elsasser symmetry is correspondingly broken, and we expect the system to have net residual energy.  This residual energy lies in the particular solution, as we will see in Section~\ref{sec:partsym}.

\subsection{Three Mode Nonresonant Interaction}\label{sec:threemode}

As a result of secular terms that grow in time and not space due to a choice of non-equilibrium initial conditions, we expect subsequent interactions involving these terms to lead to a time-dependent nonlinear drive.  This time dependence can then break the symmetry of the particular solution.  To explore this, it is necessary to calculate the particular solution response to said nonlinear drive.  We begin with a simple model for a single term in that drive:

\begin{equation}
\zeta_{a+}\zeta_{b-}=\frac{f_c(t)}{k_{a\perp}k_{b\perp}}\cos{(k_{c\parallel}z-\omega_c{t}+\theta_c)}\label{eqn:zeta_ab_onemode}
\end{equation}

\noindent which arises from the nonlinear interaction of modes ``a'' and ``b'' in the system, producing quasimode ``c.''  Here, $\omega_c$ and $k_{c\parallel}$ are the frequency and parallel wavenumber of the nonlinear drive, respectively.  The interaction is nonresonant with $\omega_c \neq \mp k_{c\parallel} V_A$.  Normalization to the perpendicular wavenumbers of modes ``a'' and ``b'' is based on Eq.~\ref{eqn:zetadef} and ensures that $f_c(t)$, which depends on a product of the Elsasser amplitudes of the two primary modes, has units of velocity squared.  $f_c(t)$ is a function of $t$ in order to capture the variation of the mode amplitude in time.  As mentioned in Section~\ref{sec:homosec}, this time dependence breaks Elsasser symmetry, and we wish to examine how this leads to residual energy generation in the particular solution.  Note that $f_c(t)$ and the phase factor $\theta_c$ can depend on the perpendicular coordinates; we assume that this dependence takes a form that allows Eqs.~\ref{eqn:k_op_f} and \ref{eqn:k_op_m} to be valid.  Fourier transforming the nonlinear drive given by $\zeta_{a+}\zeta_{b-}$:

\begin{equation}
\breve{\zeta}_{a+} \circledast \breve{\zeta}_{b-} = \frac{\sqrt{2\pi}}{2k_{a\perp}k_{b\perp}}\left[\breve{f}_c(\omega-\omega_c)\delta(k_\parallel-k_{c\parallel})e^{i\theta_c}+\breve{f}_c(\omega+\omega_c)\delta(k_\parallel+k_{c\parallel})e^{-i\theta_c}\right]\label{eqn:zeta_ab_onemode_k}
\end{equation}

\noindent here $\breve{f}_c$ represents the Fourier transform of the function $f_c$ and $\delta$ is the Dirac delta function.  Plugging this form into the first line of Eq.~\ref{eqn:ivalsolp}, we integrate over $k_\parallel$ and use Eqs.~\ref{eqn:op_p}, \ref{eqn:k_op_f}, and \ref{eqn:k_op_m} to obtain:

\begin{equation}
\begin{aligned}
\left(\zeta_{c+} \pm \zeta_{c-}\right)_{particular} = \frac{-i{\bf\hat{z}}}{\sqrt{2\pi}} \cdot \left(\bf{\hat{k}_{a\perp}} \times \bf{\hat{k}_{b\perp}}\right) \left[ \int_{-\infty}^\infty p_\pm(0,\Delta\omega+\omega_c, k_{c\parallel}) \breve{f}_c(\Delta\omega) e^{i(k_{c\parallel}{z}-(\Delta\omega+\omega_c)t+\theta_c)}d\Delta\omega \right.\\
\left. +  \int_{-\infty}^\infty p_\pm(0,\Delta\omega-\omega_c, -k_{c\parallel}) \breve{f}_c(\Delta\omega) e^{i(-k_{c\parallel}{z}-(\Delta\omega-\omega_c)t-\theta_c)} d\Delta\omega \right] \label{eqn:psolintc}
\end{aligned}
\end{equation}

\noindent where we have also performed a substitution of variables $\Delta\omega=\omega-\omega_c$ in the first integral and $\Delta\omega=\omega+\omega_c$ in the second integral.  The parallel response function $p_\pm(\ell,\omega,k_\parallel)$ captures factors resulting from the integration over the parallel coordinate $\phi'_\mp$ in Eq.~\ref{eqn:ivalsetup}.  This function depends on an integer argument $\ell$ and is defined as:

\begin{equation}
p_\pm(\ell,\omega,k_\parallel) = \frac{1}{4} \left[\frac{-1}{(\omega+k_{\parallel}{V_A})^{\ell+1}} \pm \frac{1}{(\omega-k_{\parallel}{V_A})^{\ell+1}}+\frac{k_{a\perp}^2-k_{b\perp}^2}{k_{c\perp}^2}\left(\frac{-1}{(\omega+k_{\parallel}{V_A})^{\ell+1}} \mp \frac{1}{(\omega-k_{\parallel}{V_A})^{\ell+1}}\right)\right]\label{eqn:pfn}
\end{equation}

Since we have assumed that the interactions are nonresonant, the $\omega \pm k_{\parallel}V_A$ terms in the denominator of $p_\pm$ must be finite.  From the form of Eq.~\ref{eqn:psolintc}, this means that $\breve{f}_c(\omega)$ is nonzero only in a frequency window near zero; for each frequency $\Delta\omega$ contained within this window, $|\Delta\omega| < |\omega_c \pm k_{c\parallel}V_A|$ must be satisfied to ensure that the frequency broadening due to $f_c(t)$ does not lead to a resonant interaction.  If $\omega_c$ is close to a resonant frequency $\pm k_{c\parallel}V_A$, then $|\Delta\omega|$ must always be much smaller than both $|\omega_c|$ and $|k_{c\parallel}V_A|$.   This corresponds to the case of weakly interacting modes ($\tau_A \ll \tau_{nl}$) where the amplitude variation described by $f_c(t)$ is on a much slower timescale than the sinusoidal dependence in Eq.~\ref{eqn:zeta_ab_onemode}.  Alternatively, $\omega_c$ may be far off resonance, in which case the timescale for amplitude variation described by $f_c(t)$ may be of the same order as the sinusoidal dependence such that $\tau_A \sim \tau_{nl}$.  With both these physical situations as options, we use the requirement $|\Delta\omega| < |\omega_c \pm k_{c\parallel}V_A|$ to invoke the following Taylor expansion:

\begin{equation}
\frac{1}{\Delta\omega+\omega_c \pm k_{c\parallel}V_A} = \sum_{n=0}^\infty \frac{(-\Delta\omega)^n}{\left(\omega_c \pm k_{c\parallel}V_A\right)^{n+1}}\label{eqn:tayloromega}
\end{equation}

The integral in Eq.~\ref{eqn:psolintc} can now be performed by noting that:

\begin{equation}
\frac{1}{\sqrt{2\pi}}\int_{-\infty}^\infty (-\Delta\omega)^n \breve{f}_c(\Delta\omega) e^{-i\Delta\omega{t}} d\Delta\omega = (-i)^n f_c^{(n)}(t)\label{eqn:ftdeltaomega}
\end{equation}

\noindent where the notation $f_c^{(n)}(t)$ indicates the $n$th derivative of the function $f_c(t)$.  If $f_c(t)$ is an analytic function, it can be represented by a convergent power series in the neighborhood of a time $t=t_0$.  As will be discussed in the final paragraph of Section~\ref{sec:concl}, such a series expansion for $f_c(t)$ in powers of $t$ naturally arises due to the form of nonlinear drive that is a consequence of the secular solution.  We therefore approximate $f_c(t)$ here as the polynomial ${\sum_{n=0}^m \frac{1}{n!} \left[f_c^{(n)}(t_0)\right] (t-t_0)^n}$, where the series has been truncated at $n$ equal to an integer $m$, beyond which subsequent terms are negligible due to series convergence.  This truncation is equivalent to the assumption $f_c^{(n)}(t) \equiv 0$ for $n>m$.  Under this assumption, the series in Eq.~\ref{eqn:tayloromega} will converge because only terms in this series with $n \leq m$ will be associated with a nonzero derivative of $f_c(t)$ via the integration step in Eq.~\ref{eqn:ftdeltaomega}.  Consistent with this convergence, we expect the Fourier transform of the polynomial approximation of $f_c^{(n)}(t)$ to satisfy $|\Delta\omega| < |\omega_c \pm k_{c\parallel}V_A|$.   To see this, note that the Fourier transform of $t^n$ is proportional to the $n$th derivative of the Dirac delta function.  The polynomial approximation will therefore Fourier transform into a sum of a finite number of these delta function derivatives, which by definition will be nonzero only in a very narrow frequency range near zero.  Therefore, we can use a polynomial approximation of $f_c(t)$ in order to make the present derivation valid in any small time window in which $f_c(t)$ is an analytic function.  This procedure corresponds to limiting the time window over which $f_c(t)$ is allowed to vary, which ensures that there is no frequency broadening of the nonlinear drive leading to a resonant interaction.

Following the Taylor expansion (Eq.~\ref{eqn:tayloromega}) and integration (Eq.~\ref{eqn:ftdeltaomega}), Eq.~\ref{eqn:psolintc} may be written as:

\begin{equation}
\begin{aligned}
\left(\zeta_{c+} \pm \zeta_{c-}\right)_{particular} = {\bf\hat{z}} \cdot \left(\bf{\hat{k}_{a\perp}} \times \bf{\hat{k}_{b\perp}}\right) \sum_{n=0}^\infty (-i)^{(n+1)} f_c^{(n)}(t) \left[p_\pm(n,\omega_c,k_{c\parallel})e^{i(k_{c\parallel}z-\omega_c{t}+\theta_c)} \right. \\
\left. + p_\pm(n,-\omega_c,-k_{c\parallel})e^{i(-k_{c\parallel}z+\omega_c{t}-\theta_c)} \right]
\end{aligned}
\end{equation}

Or equivalently using sine and cosine functions:

\begin{equation}
\begin{aligned}
\left(\zeta_{c+} \pm \zeta_{c-}\right)_{particular} = 2{\bf\hat{z}} \cdot \left(\bf{\hat{k}_{a\perp}} \times \bf{\hat{k}_{b\perp}}\right) \sum_{n=0}^\infty (-1)^n \left[ f_c^{(2n)}(t) p_\pm(2n,\omega_c,k_{c\parallel})\sin{\left(k_{c\parallel}z-\omega_c{t}+\theta_c\right)} \right. \\
\left. - f_c^{(2n+1)}(t) p_\pm(2n+1,\omega_c,k_{c\parallel})\cos{\left(k_{c\parallel}z-\omega_c{t}+\theta_c\right)} \right]\label{eqn:psolc}
\end{aligned}
\end{equation}

Eq.~\ref{eqn:psolc} splits modes ``c'' in the particular solution into two parts, one involving even-numbered derivatives of the $f_c(t)$ function and the sine of the phase argument, and one involving odd-numbered derivatives and the cosine of the phase argument.  We can see from this equation along with Eqs.~\ref{eqn:ufromzeta} and \ref{eqn:bfromzeta} that $\delta{v_{c\perp}}$ is proportional to a sum involving $p_+$ functions while $\delta{b_{c\perp}}$ is proportional to a sum involving $p_-$ functions.  Invoking the scale-locality assumption discussed in Section~\ref{sec:homosec}, the first two terms in Eq.~\ref{eqn:pfn} are likely to dominate the last two in a case where there are many interacting modes.  For $n=0$ (no time varying amplitudes), this means that the velocity fluctuations for each term in the particular solution will be approximately proportional to $k_{\parallel}$ while the magnetic fluctuations will be approximately proportional to $\omega$.  This result is consistent with our inferences from the physical arguments used to obtain Eq.~\ref{eqn:rA}.  These properties will be important for the discussion of the relationship between Elsasser symmetry and residual energy in the next section.

\subsection{Symmetry Breaking of the Particular Solution}\label{sec:partsym}

The form we used for $\zeta_{a+}\zeta_{b-}$ in Section~\ref{sec:threemode} is not Elsasser symmetric, even if $f_c(t)$ is set to a constant value.  We can, however, construct an illustrative example of a nonlinear drive where the symmetry may hold:

\begin{equation}
\zeta_+\zeta_- = \sum_{c} \frac{f_c(t)}{k_{a\perp}k_{b\perp}}\left[\cos\left(k_{c\parallel}z-\omega_c{t}+\theta_c\right) - \cos\left(\frac{-\omega_c}{V_A}z+k_{c\parallel}V_A{t}+\theta_c\right)\right]\label{eqn:quasisystem}
\end{equation}

\noindent where the sum is over all nonresonant modes ``c'' resulting from the coupling of all mode pairs ``a'' and ``b.''  Eq.~\ref{eqn:quasisystem} is designed following the physical arguments in the second to last paragraph of Section~\ref{sec:esym} such that modes with $|\omega_c| < |k_{c\parallel}V_A|$ and modes with $|\omega_c| > |k_{c\parallel}V_A|$ will be driven in the system in equal proportion.  If $f_c(t)$ is set to a constant, Eq.~\ref{eqn:quasisystem} is symmetric under the simultaneous variable interchange $\zeta_- \rightarrow -\zeta_-$ and $z \rightleftarrows V_A{t}$ (both sides pick up a negative sign).  However, a system consisting only of constant amplitude quasimodes will be difficult to construct because the large number of modes present is likely to include a combination that produces a resonant interaction; given our initial conditions, this leads to time-dependent mode amplitudes and non-constant $f_c(t)$.  In this context, we must consider non-constant $f_c(t)$ and should view Eq.~\ref{eqn:quasisystem} as the portion of the nonlinear drive that produces the particular solution.  For this form of $\zeta_+\zeta_-$ we can use Eq.~\ref{eqn:psolc} to write the particular solution as:

\begin{equation}
\begin{aligned}
\left(\zeta_+ \pm \zeta_-\right)_{particular} = \sum_c 2{\bf\hat{z}} \cdot \left(\bf{\hat{k}_{a\perp}} \times \bf{\hat{k}_{b\perp}}\right) \sum_{n=0}^\infty (-1)^n \biggl\{ f_c^{(2n)}(t) \biggl[p_\pm(2n,\omega_c,k_{c\parallel})\sin{\left(k_{c\parallel}z-\omega_c{t}+\theta_c\right)} \biggr.\biggr. \\
\biggl. +p_\mp\left(2n,\omega_c,k_{c\parallel}\right)\sin{\left(\frac{-\omega_c}{V_A}z+k_{c\parallel}V_A{t}+\theta_c\right)}\biggr]\\
- f_c^{(2n+1)}(t) p_\pm(2n+1,\omega_c,k_{c\parallel})\biggl[\cos{\left(k_{c\parallel}z-\omega_c{t}+\theta_c\right)}\biggr.\\
\biggl.\biggl. -\cos{\left(\frac{-\omega_c}{V_A}z+k_{c\parallel}V_A{t}+\theta_c\right)}\biggr] \biggr\}\label{eqn:psolsym}
\end{aligned}
\end{equation}

\noindent where we have used the identities i) $p_\pm\left(2n,-k_{c\parallel}V_A,-\frac{\omega_c}{V_A}\right) = - p_\mp\left(2n,\omega_c,k_{c\parallel}\right)$ and ii) $p_\pm\left(2n+1,-k_{c\parallel}V_A,-\frac{\omega_c}{V_A}\right)=p_\pm\left(2n+1,\omega_c,k_{c\parallel}\right)$; these can be proven by closely examining the form of $p_\pm$ in Eq.~\ref{eqn:pfn}.  Recall that the $p_+$ ($p_-$) function appears as the amplitude of the velocity (magnetic) fluctuations associated with a single mode in Eq.~\ref{eqn:psolc}; thus, the first identity expresses the fact that the magnetic and velocity fluctuations of the mode interchange when $z \rightleftarrows V_A{t}$.  Identity (i) is therefore both an mathematical expression of Elsasser symmetry and a generalization of Eq.~\ref{eqn:rA} to the case where the interactions may not be scale-local.  Note, however, that when $f_c^{(2n+1)}(t) \neq 0$ for any integer $n$, identity (ii) must also be considered, and this portion of the particular solution is not Elsasser symmetric because the associated magnetic and velocity fluctuations do not swap when $z \rightleftarrows V_A{t}$.

It is clear from this physical argument and the form of Eq.~\ref{eqn:psolsym} that the particular solution is Elsasser symmetric if and only if $f_c(t)$ is set to a constant.  We can see this as follows: In this case, only the first two lines are nonzero, and only for $n=0$.  The second line has the same sinusoidal dependence as the first with $z \rightleftarrows V_A{t}$, and the $p_\pm$ functions on each line swap under the $\zeta_- \rightarrow -\zeta_-$ negation.  Thus we are left with the same two terms under the simultaneous variable interchange.  But once we account for the expected variation $f_c(t)$ with $t$, both this added time dependence and the terms on the third and fourth lines may break the symmetry.

Consistent with this, there is no net residual energy generation for constant $f_c(t)$; the root mean square of both sinusoidal terms will be $1/\sqrt{2}$, and $<\delta{v_\perp}^2>$ and $<\delta{b_\perp}^2>$ will both be proportional to $p_+(0,\omega_c,k_{c\parallel})^2+p_-(0,\omega_c,k_{c\parallel})^2$.  However, when $f_c(t)$ is not constant in time, the third and fourth terms may become important, with $p_+(2n+1,\omega_c,k_{c\parallel})$ contributing to the velocity fluctuations and $p_-(2n+1,\omega_c,k_{c\parallel})$ to the magnetic fluctuations.

Assuming scale locality, $|p_-(2n+1,\omega_c,k_{c\parallel})|>|p_+(2n+1,\omega_c,k_{c\parallel})|$, which may be seen by noting that the denominators of the important first two terms in Eq.~\ref{eqn:pfn} are always positive for $\ell=2n+1$.  This means that for this part of the particular solution, we expect the energy in the magnetic fluctuations to be larger than the energy in the velocity fluctuations.  This suggests that once secularly growing modes $\sim t$ are generated, then subsequent interactions will involve a time-dependent nonlinear drive and lead to the generation of net-negative residual energy.  Even though the key physics that can break Elsasser symmetry of a system is due to the presence of the secular modes (which grow in time, not in space), the residual energy is held in the subsequently generated Alfv{\'e}nic quasimodes of the particular solution.

If we instead consider the alternate case of a boundary value problem with a condition specified at $z=0$, secular modes will grow in the parallel spatial direction rather than time (as discussed in Section~\ref{sec:homosec}).  We can model this by using $f_c(z)$ in lieu of $f_c(t)$ in our nonlinear drive.  The result of the calculation in Section~\ref{sec:threemode} will be extremely similar with the roles of $\omega$ and $k_\parallel V_A$ swapped.  In Eq.~\ref{eqn:psolsym} this swap leads to $p_\pm(2n+1, \omega_c, k_{c\parallel}) \rightarrow p_\mp(2n+1, \omega_c, k_{c\parallel})$ in the third and fourth lines.  In other words, normal modes growing in the parallel direction $\sim z$ lead to a space-dependent nonlinear drive that produces quasimodes with net-positive residual energy.  We note here that this alternate situation is somewhat unphysical, as the modes growing in space would need to be fully developed at all times, including $t=0$; however, most real systems are not perfectly described by an initial value problem either.  This exercise instead serves to illustrate how initial and boundary conditions can break Elsasser symmetry, leading to net residual energy generation; we will explore possible implications for real solar wind turbulence in the next section.

\section{Conclusions}\label{sec:concl}

We have identified a new symmetry of reduced MHD: the governing equations are symmetric under the simultaneous interchange of variables ${\bf \delta{v}} \rightleftarrows {\bf \delta{b}}$ and $z \rightleftarrows V_A{t}$ (or ${\bf \delta{v}} \rightleftarrows -{\bf \delta{b}}$ and $z \rightleftarrows -V_A{t}$), and systems that break this symmetry have net residual energy.  We demonstrate that a non-equilibrium initial condition breaks the symmetry in a way that preferentially leads to net-negative residual energy generation in scale-local interactions.  This potentially opens up a new area of magnetized plasma turbulence research, as further exploration of the new Elsasser symmetry as well as the addition of quasimodes to existing theories may yield new insights.  Several examples are discussed below.

Our analysis of the reduced MHD equations with an initial condition applied that is not an exact solution at $t=0$ contains both a particular solution at the frequency of the nonlinear drive and a homogeneous solution at the frequency of the associated normal mode. The particular solution that arises from a constant amplitude nonlinear drive may contain either positive or negative residual energy, depending on the relationship between the phase speed of each driven secondary mode and the phase speed of an Alfv{\'e}n normal mode. At the resonance where the nonlinear drive matches the normal mode frequency, secularly growing Alfv{\'e}n normal modes are produced. Due to the chosen initial conditions, these modes grow in time and not in space, breaking Elsasser symmetry. Subsequent scale-local interactions then involve a time-dependent nonlinear drive, which preferentially produces Alfv{\'e}nic quasimodes with negative residual energy.  Broken Elsasser symmetry is also evident in $\omega=k_\parallel{V_A}=0$ stationary 2D normal modes produced by scale-local interactions, as the chosen initial condition requires these modes to be purely magnetic.  By contrast, in the equivalent boundary value problem Elsasser symmetry is broken in the opposite way, leading to net-positive residual energy generation by scale-local interactions.

Our results are consistent with prior work that shows that Alfv{\'e}nic quasimodes resulting from nonlinear interactions can contain either positive or negative residual energy \citep{howes13}; this physics has also recently been corroborated in laboratory experiments \citetext{M.~Abler, 2024}.  These quasimodes are not freely propagating Alfv{\'e}n waves but retain some Alfv{\'e}n wave properties such as incompressability and a high degree of correlation between magnetic and velocity fluctuations, consistent with solar wind observations.  This suggests that the answer to the question of the fundamental nature of the observed solar wind fluctuations may be that the solar wind is dominated by Alfv{\'e}nic quasimodes that exist only in the presence of the nonlinear driving terms.

Also consistent with the physical picture presented in this manuscript, net-negative residual energy is commonly observed in prior simulation studies.  Many of these studies e.g.~\citep{oughton94, muller05, bigot08, boldyrev09, mininni09} initialize modes with equal amounts of energy in magnetic and velocity fluctuations in a periodic box, a physical situation that can be represented by the initial value problem considered in Section~\ref{sec:ival} of this manuscript that leads to net-negative residual energy generation.  As explained at the end of Section~\ref{sec:partsym}, the corresponding net-positive residual energy case contains modes that grow in space which must be fully developed at $t=0$; the impracticality of setting this up in a numerical simulation may explain why net-positive residual energy systems are not typically reported in the literature.  While these general comparisons with existing simulations are promising, more detailed theory/simulation comparisons of the early time evolution for more specific initial and boundary conditions (as done in \citet{nielson13}) are planned as a benchmark for our theoretical results.  We recently conducted such a comparison for the case of two overlapping Alfv{\'e}n waves in a periodic box and found that, as in \citet{nielson13}, the essential RMHD dynamics are the same in both the calculation and simulation; this will be presented as part of a future publication.

Our conclusion that net residual energy can be generated by a symmetry breaking in reduced MHD provides some clues that may be applicable to real turbulent systems such as the solar wind, where net-negative residual energy is also observed.  For example, our association of net-negative residual energy with an initial value problem at $t=0$ and net-positive residual energy with the equivalent boundary value problem at $z=0$ (Sections~\ref{sec:homosec} and \ref{sec:partsym}) suggests that the real solar wind case is closer to the former than the latter.  The fact that the initial condition considered in this manuscript is not an exact solution to the reduced MHD equations may also have an analogue in the solar wind.  For example, if at its origin at the solar corona, a parcel of solar wind plasma already contains a small ${\bf z^+}$ component in addition to the dominant ${\bf z^-}$ component (where ${\bf z^-}$ normal modes propagate away from the sun), an initial state that is not an exact solution could be generated.  Alternately, the expansion of the solar wind \citep{meyrand23} could modify solar wind turbulence to continuously generate states that are not exact solutions to Eq.~\ref{eqn:epoteq}, and negative residual energy could be generated upon subsequent nonlinear evolution of the plasma parcel.  A third possibility in which the solar wind consists almost entirely of ${\bf z^-}$ fluctuations at its origin and the minority ${\bf z^+}$ fluctuations are primarily produced by reflection at the Alfv{\'e}n critical point \citep{chandran09} would require initial and boundary conditions that are not considered by the present manuscript.  Therefore, an interesting area for future work is an extension of the present calculation to these and other sets of assumptions.  In the Fourier decomposition considered in this manuscript, nonlinear terms are simultaneously present at every point in space; a different physical situation with a spatially varying nonlinear drive can be used to test how Alfv{\'e}nic quasimodes change character as they enter a region where this varying nonlinear drive supports a different set of quasimodes.  Possible choices for future work include initial and boundary conditions similar to laboratory experiments \citetext{\citealp{drake16}; M.~Abler, 2024; C.~H.~K.~Chen, 2024} or the case of two interacting Alfv{\'e}nic wavepackets simulated by \citet{verniero18}.  Since real turbulent systems may also include a subdominant compressive component, resistivity, viscosity and/or kinetic physics, examining how these physical effects may break Elsasser symmetry is also a promising future direction.  We might expect physics that has been previously shown to be of negligible importance to the turbulent cascade in the solar wind (e.g.~compressive fluctuations \citep{goldreich95,schekochihin19}) to break Elsasser symmetry differently than other effects (e.g.~kinetic effects at ion scales \citep{groselj19}) that play a more central role.

Other fundamental questions about Elsasser symmetry are also ripe for follow-on work.  For example, while we demonstrated in Section~\ref{sec:esym} that an Elsasser symmetric system cannot have net residual energy, we have not proven the converse.  As discussed in that section, modes with positive and negative residual energy exist in equal proportion in an system with no net residual energy, and under the simultaneous variable interchange each mode will transform into one of the opposite type.  However, each type of mode will only exactly transform into the other, as required for Elsasser symmetry, if there exists a supplementary coordinate transformation that can eliminate any discrepancy in the sinusoidal phases.  We postulate that such a transformation exists for any zero net residual energy system, but proof of this conjecture is left to future work. 

Future work may also focus on the closure issues associated with the secular term.  Formally, the right side of Eq.~\ref{eqn:ivalsolc} is no longer a Fourier decomposition due to the factor of $t$ in the secular term.  These secular modes lead to a nonlinear operator proportional to $t$ times a sinusoidal dependence, which at resonance leads to terms of power $t^2$ in the secular solution; the continuation of this logic will clearly cause even higher powers of $t$ and hence a closure problem.  To avoid this problem in Section~\ref{sec:threemode}, we modeled the nonlinear drive that results from this time variation by the function $f_c(t)$; a closed form solution will therefore yield the form for $f_c(t)$ for each mode in the system's nonlinear drive.  A possible way to attack this is to borrow multiple scale methods from WT theory in which the mode amplitudes are taken to be slowly varying compared to the fluctuating component \citep{galtier00}.  Insights from this exercise, which will essentially add Alfv{\'e}nic quasimodes to WT theory, may then motivate any further mathematical development required to obtain a closed-form solution to the more general case.  This assumes of course that such a solution is possible, which is a question left to future work.

\section*{Acknowledgments}
The four co-authors are listed alphabetically to acknowledge contributions of approximately equal importance, consisting of many fruitful discussions and helpful comments that substantially improved this manuscript.  G.~Howes provided very helpful comments on an initial draft of a related paper.  Discussions with Bindesh Tripathi regarding multiple scale methods are also acknowledged.  S.~Dorfman was supported by NASA grant 80NSSC18K1235 and DOE Grant DE-SC0021291.  M.~Abler was supported by DOE grant DE-SC0023326.  S.~Boldyrev was supported by the U.S. Department of Energy, Office of Science, Office of Fusion Energy Sciences under award number DE-SC0024362, and by the NSF grant PHY-2010098.  C.~H.~K.~Chen was supported by UKRI Future Leaders Fellowship MR/W007657/1 and STFC Consolidated Grants ST/T00018X/1 and ST/X000974/1.  S.~Greess was also supported by UKRI Future Leaders Fellowship MR/W007657/1.
\bibliographystyle{aasjournal}



\end{document}